\documentclass{article}

\usepackage{arxiv}

\usepackage[utf8]{inputenc} % allow utf-8 input
\usepackage[T1]{fontenc}    % use 8-bit T1 fonts
\usepackage{hyperref}       % hyperlinks
\usepackage{url}            % simple URL typesetting
\usepackage{booktabs}       % professional-quality tables
\usepackage{amsfonts}       % blackboard math symbols
\usepackage{nicefrac}       % compact symbols for 1/2, etc.
\usepackage{microtype}      % microtypography
\usepackage{lipsum}
\usepackage{amsmath}
\usepackage{graphicx}
\usepackage{nicefrac}
\usepackage[lofdepth,lotdepth]{subfig}
\title{A critique of the Mean Field Approximation in preferential attachment networks}

\author{
  Matthijs~Ruijgrok \\
  Department of Mathematics\\
  Utrecht University\\
  P.O. Box 80010, 3508 TA Utrecht \\
  The Netherlands \\
  \texttt{m.ruijgrok@uu.nl} \\
}

\begin{document}
\maketitle

\begin{abstract}
The Mean Field Approximation (MFA), or continuum method, is often used in courses on Networks to derive the degree distribution of preferential attachment networks. This method is simple and the outcome is close to the correct answer. However, this paper shows that the method is flawed in several aspects, leading to unresolvable contradictions. More importantly, the MFA is not explicitly derived from a mathematical model. An analysis of the implied model shows that it makes an approximation which is far from the truth and another one which can not be motivated in general. The success of the MFA for preferential attachment networks is therefore accidental and the method is not suitable for teaching undergraduates.  
\end{abstract}

% keywords can be removed

\section{Introduction} 
The birth of Network Science occurred somewhere in the late nineties. One of its foundational papers is the article by Barabási and Albert \cite{BarAlb} on preferential attachment networks, which has become the most cited paper in the discipline with currently more than 36000 citations. The paper introduced an analytical method to calculate the node degree distribution of preferential attachment networks, which are now known as Barabási-Albert (BA) networks. This method is called the Mean Field Approximation (MFA) or continuum method.  The result of the MFA is close to correct, but not exactly. The MFA has since been reproduced in many textbooks, see for example \cite{Bar}, \cite{Barr} and \cite{Cald} and in (online) course notes, see  \cite{MIT} and \cite{Stanf} (specifically \cite{Stanf2}) .  \\

In this paper, I will analyse the MFA from the point of view of mathematical modelling. The conclusion is that the method is the result of poor modelling, and its relative success is accidental. Because the MFA is so widely used in courses on networks, I believe it is worthwhile to point out its failings.\\\\
The motivation for writing this paper is my experience with teaching the MFA in an undergraduate course on networks. I encountered two major problems with the method. First, the MFA is a sequence of steps, each of which is familiar from introductory undergraduate mathematics courses. However, none of the steps is carefully motivated, but simply applied mechanically. It obscures where the approximations are made, and what assumptions are necessary. Especially for students without much experience in the field, it gives a wrong impression of mathematical modelling. \\
The second problem with the MFA is in its details. It contains fallacies that lead to contradictions, makes arbitrary choices, confuses stochastic variables with deterministic ones, uncritically changes discrete variables to continuous ones and produces an incorrect answer.\\
This last set of problems is actually not the most serious of the two. By slightly reformulating the method, these problems can be removed, except for the not quite correct final answer. However, the reformulation reveals what modelling choices had been implicitly made. A closer look at the reconstructed full mathematical model shows that it relies on a very poor approximation, and an assumption which, although almost true for BA networks, is false in general. The modelling process can not be rehabilitated and this is the reason I believe the MFA is not a suitable method to teach to students. \\

\section{Background}
The BA network starts at time $t=1$ with one node, which has $m\geq 1$ self-loops. \footnote{This choice is made for simplicity. Starting with a network with more nodes and arbitrary connections doesn't change its properties for large $t$.} At time-step $t+1$, a new node is added to the network. The new node has $m$ outgoing links, each of which is attached to one of the $t$ existing nodes. The probability that the new node links to a certain existing node is taken to be proportional to the degree of this target node. This process is then repeated.\\
The primary quantity of interest for the BA network is its expected degree-distribution $p(k;t)$, which is the expected fraction of nodes at time $t$ that have degree $k$. Equivalently, $p(k;t)$ is the probability that a randomly and uniformly chosen node at time $t$ has degree $k$.\\
Numerical simulations in \cite{BarAlb} show that, when averaged over many realisations, $p(k;t)$ converges in time to a stationary distribution $ p(k)$. In particular, $p(k)\approx c \,k^{-\gamma}$ for large $k$, where $\gamma = 2 .9\pm 0.1$ and $c$ a constant. \\ 
The MFA predicts:
\begin{eqnarray}
\label{degBarAlb}
p(k) \approx \frac{2m^2} {k^3} \, ,
\end{eqnarray} 
for large $k$.\\
Soon after \cite{BarAlb}, Krapivsky, Redner and Leyvraz \cite{KraRedLey} and Dorogovtsev, Mendes and Samukhin \cite{DorMenSam}, simultaneously and independently, derived the approximation:
\begin{eqnarray}
\label{degmaster}
p(k) \approx \frac{2m(m+1)}{k(k+1)(k+2)} \, ,
\end{eqnarray} 
valid for all $k \geq m$.\\
For large $k$, both (\ref{degBarAlb}) and (\ref{degmaster}) have a leading term of order $k^{-3}$, although the prefactors differ. In contrast to (\ref{degBarAlb}), expression (\ref{degmaster}) defines a probability distribution, because $\sum_{k\geq m} p(k)=1$. Both 
 Krapivsky et al. and Dorogovtsev et al. use the rate equation method to arrive at the result. See Appendix A for a quick review of this method.\\
In $2001$ Bollob\'as et al. \cite{BolEtAl} proved that, when couched in a careful mathematical formulation, the limit $p(k)$ indeed exists and that the approximation (\ref{degmaster}) is exact. In the following years, research on networks quickly expanded, both in its foundations and in finding applications in many fields. Soon, textbooks were published and Network Science became  a subject of undergraduate curricula. Many authors chose to use the MFA to analyse BA networks, although in most cases these books and notes also discuss the rate equation method, and mention both result (\ref{degBarAlb}) and (\ref{degmaster}).

\section{The Mean Field Approximation}
\label{sec3}
In the BA model, time is discrete and will be denoted as $t_1=1, t_2=2, \dots$. Nodes will be identified by the moment in time they were added to the network, also known as their birth date. The degree at time $t$ of a node born at time $t_i$ is denoted $K_i(t)$, with $t \geq t_i$. For all nodes, $K_i(t_i)=m$.\\
The probability that a node $t_i$ is chosen as a target node at time $t+1$ is proportional to $K_i(t)$, specifically:
\begin{equation}
\label{bigpi}
\Pi_i(t)=\frac{K_i(t)}{\sum_{j=1}^{j=N(t)}K_j(t)}\, ,
\end{equation}
where $N(t)$ is the number of nodes in the network at time $t$.

\subsection{The procedure}
\label{method}
The following description of the MFA follows \cite{BarAlb}  and \cite{BarAlbJeo}.
\begin{itemize}
\item [(a)] Let $N(t)$ be the number of nodes at time $t$ and $S(t)=\sum_{j=1}^{j=N(t)}K_j(t)$. Obviously $N(t)=t$. Since at each time $t=1,2,\dots$ a new node is introduced which has $m$ outgoing links and the rest of the network receives $m$ incoming links, it is clear that:
\begin{align}
\label{sumdegree}
S(t)=2mt \, .
\end{align}
\item [(b)] Assume that the $K_i(t)$ and $t$ are reals, rather than integers. The rate of change of $K_i(t)$ per unit time is equal to $m \Pi_i(t)$, giving
\begin{align}
\label{propeq}
\frac{{\text{d}} K_i}{{\text{d}} t}=m \Pi_i(t)=\frac{m K_i}{S(t)}=\frac{K_i}{2t} \, , \quad K_i(t_i)=m\, , 
\end{align}
with solution
\begin{align}
\label{kit}
K_i(t)=m(\frac{t}{t_i})^{\nicefrac{1}{2}}\, .
\end{align}
\item [(c)] Let $P(k;t)$ be the probability that, at time $t$, a randomly and uniformly chosen node has degree $k$ or larger. Because there is a one-to-one correspondence between the birth date of a node and its degree, given by (\ref{kit}), this probability is equal to the probability that the birth date of this node is $t_i$ or less.  All nodes born after this date have degree smaller than $k$. Therefore,
\begin{align}
\label{Pwrong}
P(k;t)={\text{Pr}}(t_i\leq (\dfrac{m}{k})^2 t)=(\dfrac{m}{k})^2 \, .
\end{align}
The last equality follows from the fact that $t_i$ is chosen with uniform probability from the interval $[0,t]$.
\item [(d)]  The degree-distribution of the network at time $t$ is given by 
\begin{align}
\label{degMFA}
p(k;t)&=-\frac{\partial P(k;t)}{\partial k} \nonumber \\
&=\frac{2m^2}{ k^3}\, .
\end{align}
\end{itemize}

\subsection{Problems with the procedure}
\label{crit}
The above steps raise many objections, as itemized below:
\begin{itemize}
\item Since $S(t)=2mt$ is a deterministic function, it is implied that $K_i(t)$ is also a deterministic quantity, which it is not. The degree $K_i(t)$ is a stochastic variable, with a corresponding probability distribution over all its possible outcomes.
\item The equality $S(t)=2mt$ is true for integer values of $t$, but not for other values. Since no nodes or edges are added between integer times $t=n$ and $t=n+1$, the value of $S(t)$ is constant during that time. 
\item The degree $K_i(t)$ has integer values. In the MFA, $K_i(t)$ can take on non-integer values, the meaning of which is not explained.
\item The erroneous formula $S(t)=2mt$ leads to a contradiction. Take $t \in (n-1,n)$. Summing both sides of $ K'_i(t)=m \Pi_i(t)$ from $i=1$  to $i=n-1$ gives $S'(t)=m$, with solution $S(t)=mt+c$, which is obviously not the same as $2mt$.\\
\item If the assumption of $K_i(t)$ as a deterministic quantity is accepted, no further steps are needed. Define the jump points at time $t$ as
\begin{align}
\label{bark}
 \bar{k}_i(t)=m(\frac{t}{t_i})^{\nicefrac{1}{2}}\, .   
\end{align}
Picking a node at time $t$, randomly and uniformly, produces a probability distribution
\begin{align}
\label{tildep}
\Tilde{p}(k;t)=\begin{cases} \frac{1}{t} & \text{if } k=\bar{k}_i(t)\\
                      0                                    & \text{otherwise\, , }     
        \end{cases}
\end{align}
which is by definition the sought-after degree distribution.
\item Equation (\ref{Pwrong}) is not correct. Since the degrees of the nodes in the network have integer values, the true $P(k;t)$ is a piece-wise constant function of $k$, with jumps in integer values of $k$. 
 
\end{itemize}

\newpage
\subsection{A visual overview of the MFA}

\begin{figure}[thbp]
\centering
 \subfloat[Assuming that $K_i(t)$ is a deterministic quantity, the probability distribution $\Tilde{p}(k;t)$ for the degree of a randomly and uniformly chosen node at time $t$ is uniform and concentrated in $k=\bar{k}_i(t)$, $i=1,2, \dots, t$. ]{\includegraphics[width=0.45\textwidth]{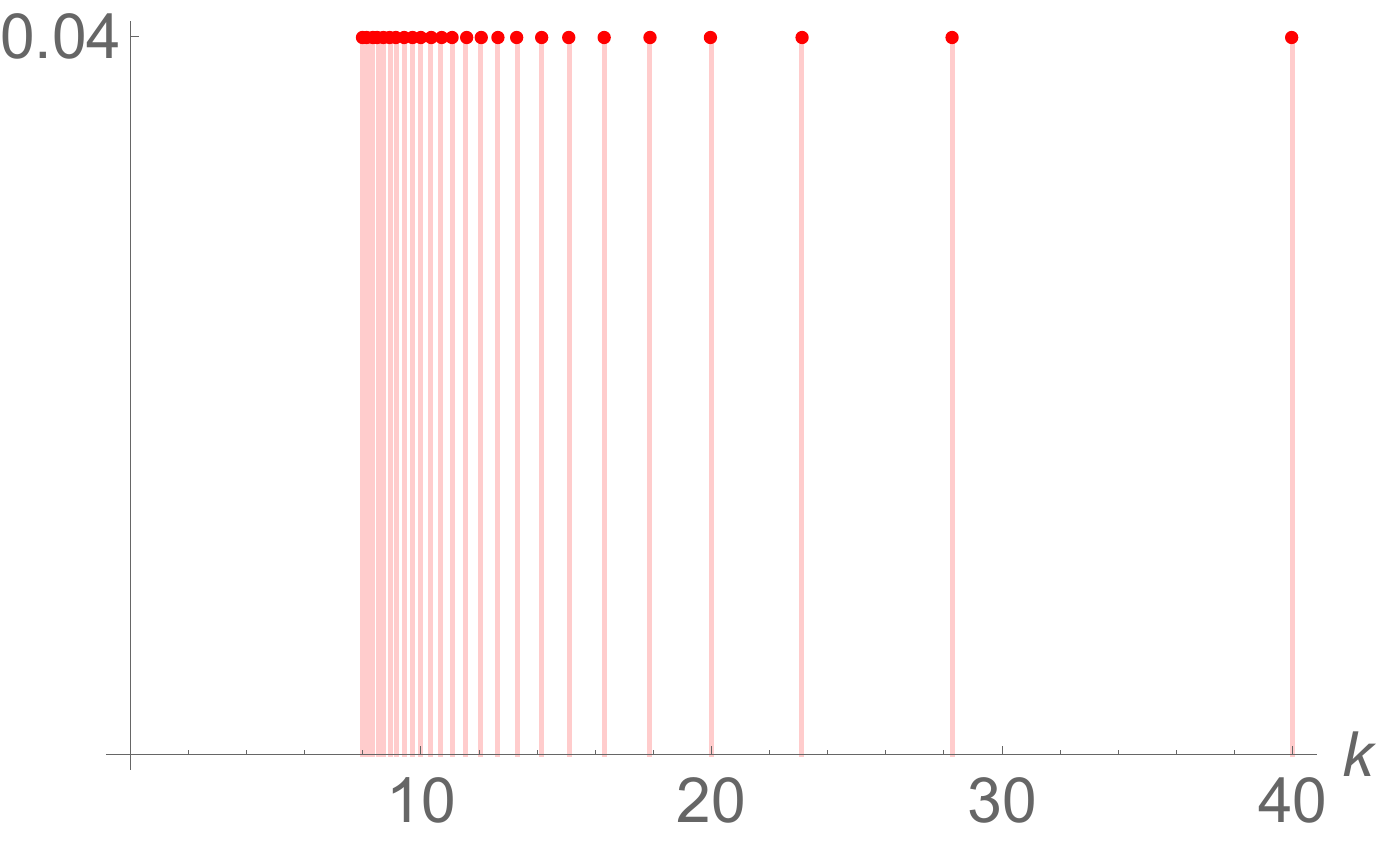}}
 \qquad
 \subfloat[$\Tilde{P}(k;t)=Pr(K_i(t)\geq k)$ (red) is the (complementary) cumulative distribution corresponding to $\Tilde{p}(k;t)$.  It is approximated by $P_{MFA}(k;t)=(m/k)^2$ (black).  ]{\includegraphics[width=0.45\textwidth]{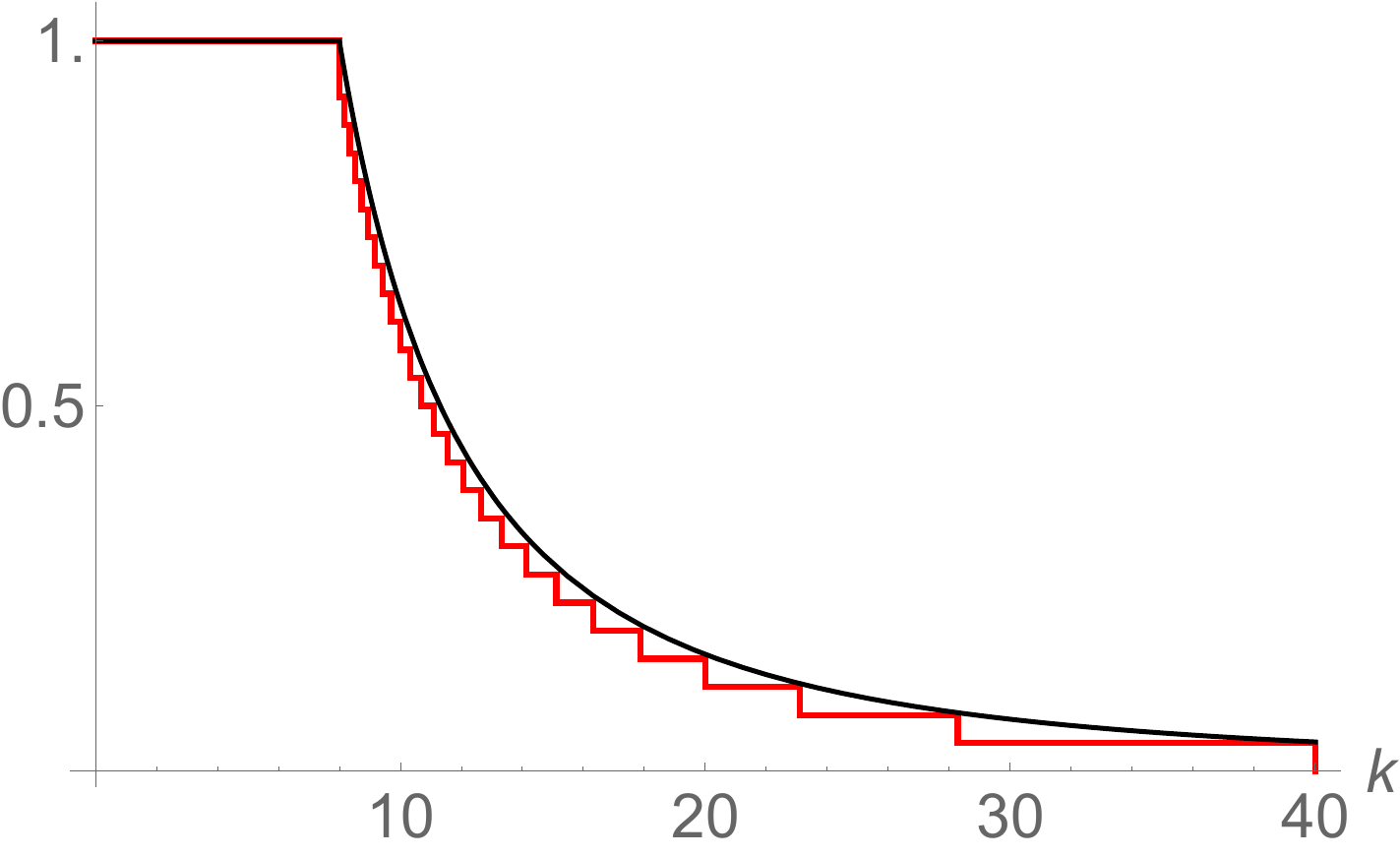}
 \label{b}}\\
 \subfloat[Define $p_{MFA}(k)=-\dfrac{\partial }{\partial k}P_{MFA}(k;t)=(2m^2)/(k^3)$ (blue). Compared with the starting distribution $\Tilde{p}(k;t)$ (red), a considerable redistribution of probability has occurred. ]{\includegraphics[width=0.45\textwidth]{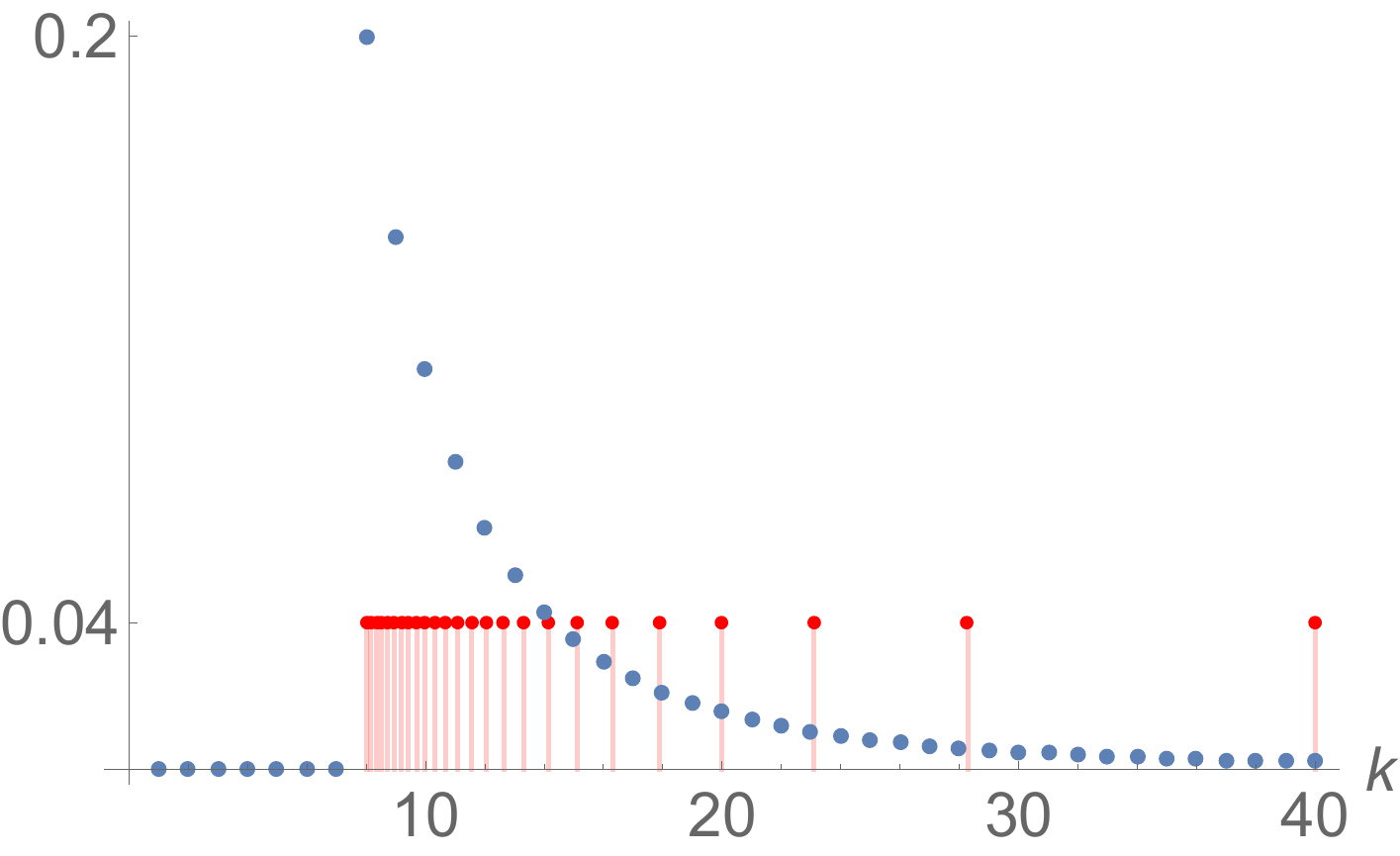}
 \label{c}}
 \qquad
 \subfloat[The cumulative distribution $\Tilde{P}(k;t)$ (red) agrees well with the complementary cumulative distribution of the true $p(k;t)$ (blue) given by (\ref{degmaster}).]{\includegraphics[width=0.45\textwidth]{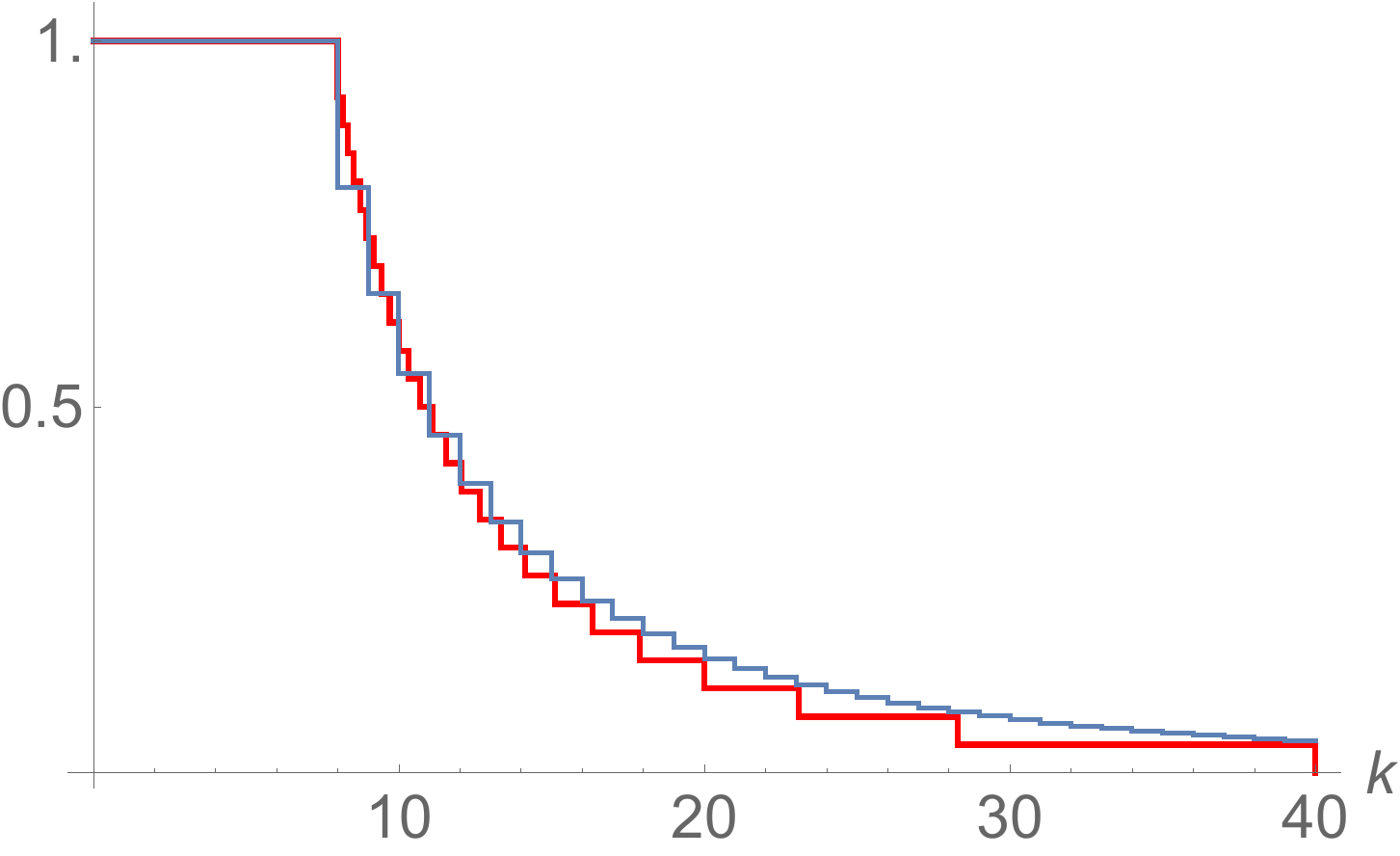}
 \label{1d}}\\
 \caption{The MFA method in pictures. In all graphs, $t=25$ and $m=8$.}
\label{MFAfig}
\end{figure}
Apart from the errors mentioned in \ref{crit}, the procedure lacks a heuristic content, as illustrated in Figure \ref{MFAfig}. The procedure starts with the probability distribution $\Tilde{p}(k;t)$. It is clear that $\Tilde{p}(k;t)$ is not in any way close to the true $p(k;t)$, if only because the support of $\Tilde{p}(k;t)$ is the set of $\bar{k}_i(t)$ rather than the integers. Then, the corresponding cumulative distribution $\Tilde{P}(k;t)$ is approximated by $P_{MFA}(k;t)=(m/k)^2$. This expression is  differentiated with respect to $k$, and its evaluation on the integers is declared to be the sought-after solution.\\
There is no obvious reason why this redistribution of probabilities should lead to an approximation of $p(k;t)$.

Note that Figure \ref{1d} shows that $\Tilde{P}(k;t)$, although derived from incorrect assumptions and through a questionable redistribution of probability, agrees well with the complementary cumulative distribution of the correct expression (\ref{degmaster}). For large values of $k$, the error seems to be no larger than $1/t$.
\newpage
\section{The implied model}
There are two sources for the problems identified in \ref{crit}. The first is the extension of $t$ and $K_i(t)$ to continuous quantities. Since there is no assumption of anything changing in the network between two integer times, and degrees of nodes are always integer valued, these extensions are unnecessary. The second source is taking $K_i(t)$ to be deterministic. The intention of this assumption is that $K_i(t)$, as it is used in the MFA, should be seen as the expected value of the degree of node $i$ at time $t$. This becomes clear by noting that the MFA value of $K_i(t)$ is defined by the differential equation (\ref{propeq}), the right hand side of which is the amount by which the expected value of the actual $K_i(t)$ changes during one time step. Moreover, it is apparent that rather than taking $K_i(t)$ as deterministic, the MFA makes the approximation that the probability distribution for $K_i(t)$ is completely concentrated in a single point. Obviously, this single point is the expected value of $K_i(t)$.
\subsection{The mathematical model corresponding to the MFA}
\label{mathmod}
The above considerations imply the following mathematical model which follows the steps of the procedure outlined in \ref{method}, but without its contradictions. Start with the description of the model as in section \ref{sec3}. An approximation $P_{MFA}(k;t)$ of the cumulative probability function $P(k;t)$ will be calculated using the probability distributions of the $K_i(t)$. From this approximation, a probability distribution $p_{MFA}(k;t)$ can then be derived.
\begin{itemize}
\item[(a)] 
Let  $B(t)$ be the birth date of a randomly and uniformly chosen node at time $t$. Clearly, $Pr(B(t)=t_i)=1/t$, for all $i=1, 2, \ldots, t$. Then
\begin{align}
\label{cumul}
    P(k;t)&=\sum_{i=1}^{t}Pr(B(t)=t_i)Pr(K_i(t)\geq k)=\frac{1}{t}\sum_{i=1}^{t}Pr(K_i(t)\geq k) \nonumber\\
    &=\frac{1}{t}\sum_{i=1}^{t}\sum_{j=k}^{\infty} p_i(j;t)\, ,
\end{align}
where $p_i(j;t)$ is the probability that $K_i(t)=j$, at time $t$.
\item[(b)]
Now approximate the probability distribution $p_i(x;t)$ of $K_i(t)$ by the delta-distribution:
\begin{align}
\label{ptil}
\Tilde{p}_i(x;t)&=
\begin{cases} 
1\, \, \text{if}\, \, x=\langle K_i(t) \rangle \\
0 \,\,\text{otherwise }     
        \end{cases}
\end{align}
with $\langle K_i(t) \rangle$ the expected value of $K_i(t)$. 
\item[(c)] The above assumption leads to:
\begin{align*}
    \sum_{j=k}^{\infty} \Tilde{p}_i(j;t)=\begin{cases} 
1\, \, \text{if}\, \, k\leq \langle K_i(t) \rangle \\
0 \,\,\text{otherwise }     
        \end{cases}
\end{align*}
For a given value of $k$, let $i=b(k;t)$ be the smallest index such that  $k\leq \langle K_i(t) \rangle$ at time $t$. Then $P(k;t)$ is approximated by:
\begin{align*}
    \Tilde{P}(k;t)=\frac{1}{t}\sum_{i=1}^{b(k;t)}1=\frac{1}{t}b(k;t).
\end{align*}
\item[(d)]
From the definition of the model, the following recursion can be obtained:
\begin{align}
\label{recurK}
  \langle K_i(t+1) \rangle  =\langle K_i(t) \rangle+\frac{\langle K_i(t) \rangle}{2t} \, , \quad \langle K_i(t_i)\rangle=m\,.
\end{align}
The solution of (\ref{recurK}) is approximated by the following differential equation (for ease of presentation, this approximation will still be denoted $\langle K_i \rangle$):
\begin{align*}
\frac{{\text{d}} \langle K_i \rangle }{{\text{d}} t}=\frac{\langle K_i \rangle }{2t} \, , \quad \langle K_i(t_i) \rangle=m\, , 
\end{align*}
with solution
\begin{align}
\label{appK}
\langle K_i(t) \rangle=m(\frac{t}{t_i})^{\nicefrac{1}{2}}\, .
\end{align}
\item[(e)] From (\ref{appK}) follows:
\begin{align*}
    b(k;t)=\lfloor (m/k)^2 t \rfloor \, , 
\end{align*} 
and
\begin{align}
\label{Ptil}
    \Tilde{P}(k;t)=\frac{1}{t}\lfloor (\frac{m}{k})^2 t \rfloor \, .
\end{align}
\item[(f)] Make the following approximation:
\begin{align}
\label{Papp}
    P_{MFA}(k;t)=(\frac{m}{k})^2 \, . 
\end{align}
\item[(g)] The value of $p_{MFA}(k;t)$ is then defined as
\begin{align}
\label{pappc}
    p_{MFA}(k;t)=P_{MFA}(k;t)-P_{MFA}(k+1;t)\, .
\end{align}
\item[(h)] The right hand side of (\ref{pappc}) can be approximated by a derivative, giving
\begin{align*}
    p_{MFA}(k;t)=-\frac{\partial}{\partial k}(\frac{m}{k})^2=\frac{2m^2}{k^3}\, .
\end{align*}
\end{itemize}
\subsection{Problems with the model}
Although the model in \ref{mathmod} is now free of contradictions, it has many questionable aspects, listed below.
\begin{itemize}
\item [(b)] The distributions $p_i(j;t)$ are concentrated on integer values of $j$, whereas the support of $\Tilde{p}_i(x;t)$ is on the values $x=\langle K_i(t) \rangle$. Since it is not likely that these values are close to integers, $\Tilde{p}_i(x;t)$ is in no sense an approximation of $p_i(j;t)$.\\
Moreover, (\ref{ptil}) is very crude. It assumes that all the probability of $K_i(t)$ is concentrated in a single point. Put another way, this assumption says that all moments of $K_i(t)$ apart from the mean are negligible. Appendix B, which presents a method to calculate the $p_i(j;t)$, shows that this assumption is far from the truth. The distributions $p_i(j;t)$ have a considerable variance.  
\item [(c)] Given the previous remark, there is no à priori reason to believe that $\Tilde{P}(k;t)$ is close to the true $P(k;t)$.
\item [(d)] The recursion (\ref{recurK}) can be solved without recourse to the dubious method of replacing it with a differential equation. Appendix C shows that the result is still close to expression (\ref{appK}), at least when $t_i$ is not too small.
\item[(f)] The function $P_{MFA}(k;t)$ is exactly equal to $\Tilde{P}(k;t)$ for all values of $k$ such that $(\dfrac{m}{k})^2 t$ is an integer. These are the jump points $k=\bar{k}_i(t)$.\\ However, for other values of $k$, notably the integers, there is no reason to choose the particular form (\ref{Papp}) as an approximation. There are many functions that agree with $\Tilde{P}(k;t)$ in the jump points and are decreasing in $k$. Each of these different functions will lead to different values of $p_{MFA}(k;t)$. Because the distances between successive jump points is increasing without bound, $\Tilde{P}(k;t)$ simply does not provide enough data points to make a reasoned guess what the values of $P(k;t)$ might be in integer values of $k$.  
\item[(h)] This step is superfluous.
\end{itemize}
The objection against item (f) already invalidates the reasoning behind the statement that $2m^2/k^3$ is a good approximation of $p(k;t)$. \\
However, when (\ref{Ptil}) is accepted as an approximation of $P(k;t)$, a weaker result can be derived, which only uses the value of $\Tilde{P}(k;t)$ in the jump points. The average value of $\Tilde{p}(k;t)$ on the interval between two successive jump points can be calculated, recalling that $\bar{k}_{i}(t)>\bar{k}_{i+1}(t)$ and $\Tilde{P}(k;t)$ is a decreasing function of $k$:
\begin{align*}
\Tilde{p}_{av}(\bar{k}_{i}(t);t)&=\frac{1}{(\bar{k}_{i}(t)-\bar{k}_{i+1}(t))} \sum_{j\geq \bar{k}_{i+1}(t)}^{j< \bar{k}_{i}(t)}\Tilde{p}(j;t) \\
&=\frac{\Tilde{P}(\bar{k}_{i+1}(t);t)-\Tilde{P}(\bar{k}_{i}(t);t)}{(\bar{k}_{i}(t)-\bar{k}_{i+1}(t))}\, .
\end{align*}
Using expressions (\ref{Ptil}) and (\ref{bark}), and the fact that $t_{i+1}=t_i +1$, this becomes:
\begin{align}
\label{pav}
\Tilde{p}_{av}(\bar{k}_{i}(t);t)= \frac{2t_i^{\nicefrac{3}{2}}}{m t^{\nicefrac{3}{2}} }+\mathcal{O}(t_i^{-1})=\frac{2 m^2}{\bar{k}_{i}^3(t)}+\mathcal{O}(t_i^{-1}) \, .
\end{align}
Without making further assumptions, (\ref{pav}) is the strongest possible statement about the asymptotic behaviour of $p(k;t)$ for large values of $k$, using only the values of the means of the $K_i(t)$.

\section{Why does the MFA work for BA networks?}
\label{theor}
\begin{figure}[htbp]
 \centering
 \includegraphics[width=0.4\textwidth]{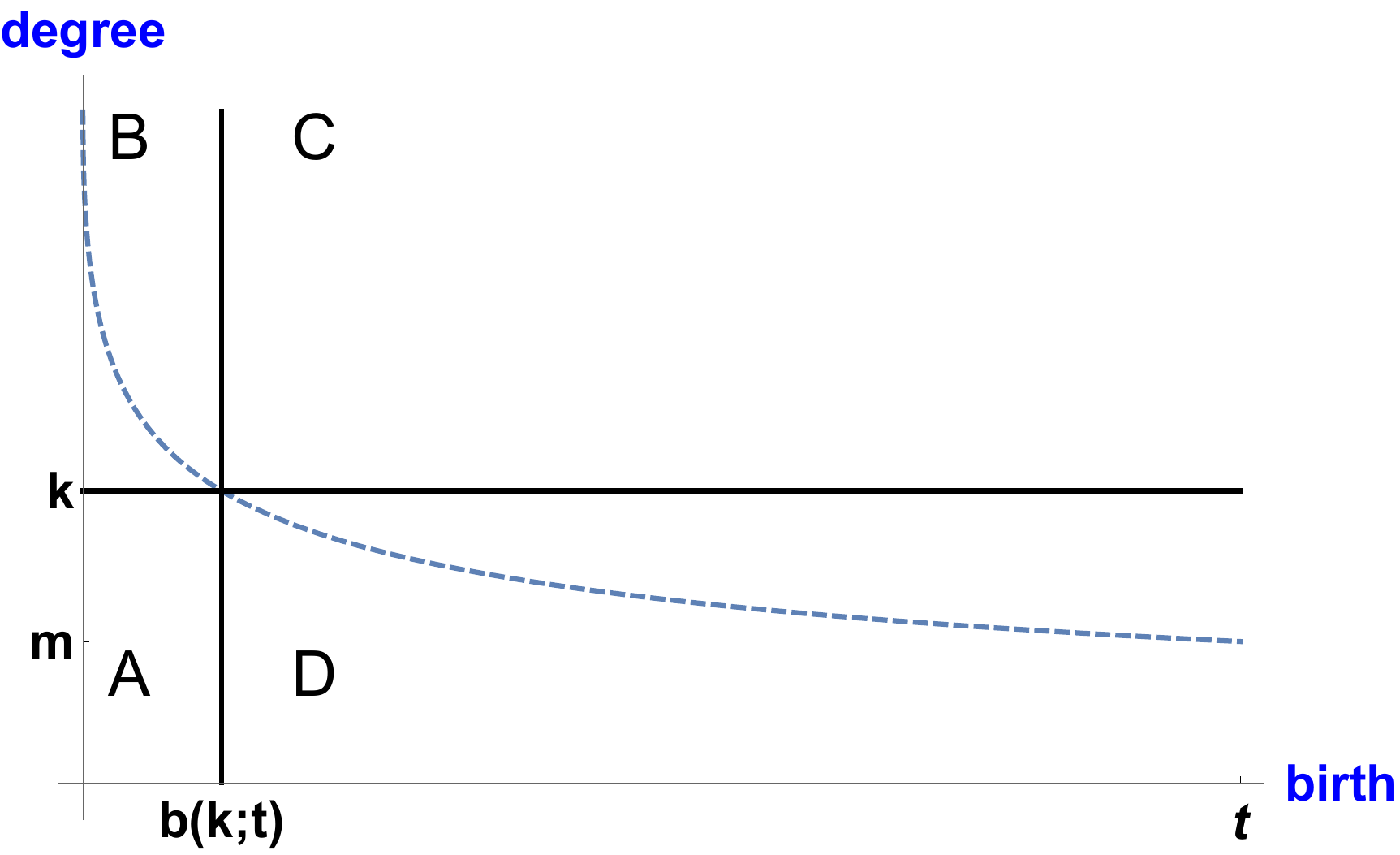}
  \caption{ A partition of the domain of $p(l,t_i)$, the probability that at time $t$ a uniformly randomly chosen node has birth date $t_i$ and degree $l$. The black horizontal line shows $l=k$, the vertical line $t_i=b(k;t)$. The dashed line is $\langle K_i \rangle$. }
  \label{partition}
\end{figure}
The MFA delivers a result which, even in its weak form (\ref{pav}), is reasonably close to the correct expression. Considering the above objections, this is an unexplained phenomenon. The answer to this conundrum is found in a closer inspection of step (c), which is the crucial part of the model. In this step, the true $P(k;t)$ is pronounced to be close to the constructed $\Tilde{P}(k;t)$. This is not supported by any heuristic, since the underlying probability distributions are very dissimilar, as illustrated in Figure \ref{MFAfig}. However, it is possible to derive a condition under which this closeness is indeed achieved. To this end, rewrite:
\begin{align}
\label{partiti}
\Tilde{P}(k;t)&=\frac{1}{t} \sum_{i=1}^{b(k;t)} 1  \nonumber \\
 &= \frac{1}{t} \sum_{i=1}^{b(k;t)}\sum_{l=m}^{\infty} p_i(l,t)\nonumber\\
 &=\frac{1}{t} (\sum_{l=m}^{k-1}\sum_{i=1}^{b(k;t)}p(l,t_i)+ \sum_{l=k}^{\infty}\sum_{i=1}^{b(k;t)}p(l,t_i)) \nonumber \\
 &=\frac{1}{t} ( \sum_{A}p(l,t_i)+\sum_{B}p(l,t_i)) \, .
\end{align}
On the other hand, $P(k,t)$ can be written as
\begin{align}
\label{partitip}
P(k;t)&= \frac{1}{t} (\sum_{l=k}^{\infty}\sum_{i=1}^{b(k;t)}p(l,t_i)+\sum_{l=k}^{\infty}\sum_{i=b(k;t)+1}^{t}p(l,t_i)) \nonumber \\
 &=\frac{1}{t} (\sum_{B}p(l,t_i)+\sum_{C}p(l,t_i)) \, .
\end{align}
See Figure \ref{partition} for an illustration of the areas $A$, $B$ and $C$.\\
Comparing (\ref{partiti}) and (\ref{partitip}) shows that $\Tilde{P}(k;t)$ is a good approximation of $P(k;t)$  if and only if the relative error
\begin{align}
\label{AC}
\frac{|\sum_{A}p(l,t_i)-\sum_{C}p(l,t_i)| }{\sum_{B}p(l,t_i)}  
\end{align} 
is small.\\
The expression $(\sum_{A}p(l,t_i))/t$ gives the probability that, at time $t$, a uniformly randomly chosen node has a birth date before $t_i=b(k;t)$ and a degree smaller than $k$, whereas $(\sum_{C}p(l,t_i))/t$ is the probability that such a random node has birth date after $b(k;t)$ and a degree larger than $k$. For the MFA to work, the error (\ref{AC}) must be small for all large values of $k$.\\
The distribution of $p(l,t_i)/t$ for the case of the BA network is shown in Figure \ref{contour}, using the results found in Appendix B. The figure is somewhat stylized, since it shows the distribution as if $t_i$ and $l$ were continuous variables. This is only done for illustration purposes.\\
For large values of $k$, the value of $b(k;t)$ is small. Then, $\sum_{A}p(l,t_i)/t$ is roughly the product of a small probability (the birth date of the chosen node is less than $t_i=b(k;t)$) and a quite substantial probability (the degree of the node, conditional on its birth date, is less than $k$). For $\sum_{C}p(l,t_i)/t$ the magnitudes of the probabilities are reversed. Apparently, the products of these probabilities in $A$ and $C$ are more or less the same size. This balancing phenomenon results from the facts that the width of the distribution $p(l,t_i)$ is  larger for small values of $t_i$ compared to larger ones and that $p(l,t_i)$ is somewhat skewed (see Figure \ref{simuls}).\\
There is, however, no reason to expect that in general the relative error (\ref{AC}) is small. Moreover, checking whether it is indeed small requires detailed knowledge of $p_i(l;t)$ beyond its expected value. If such information is available, an approximation such as the MFA becomes superfluous, since then the exact expression for $P(k,t)$ given by (\ref{partitip}) can be used.

\begin{figure}[htbp]
 \centering
 \subfloat[Contour plot of $p(l+m,t_i)/t$. Horizontally $0\leq t_i\leq 30$, vertically $l$. Blue indicates low values, red high values. The black curve is $\langle K_i(t) \rangle$ as a function of $t_i$.  ]{\includegraphics[width=0.3\textwidth]{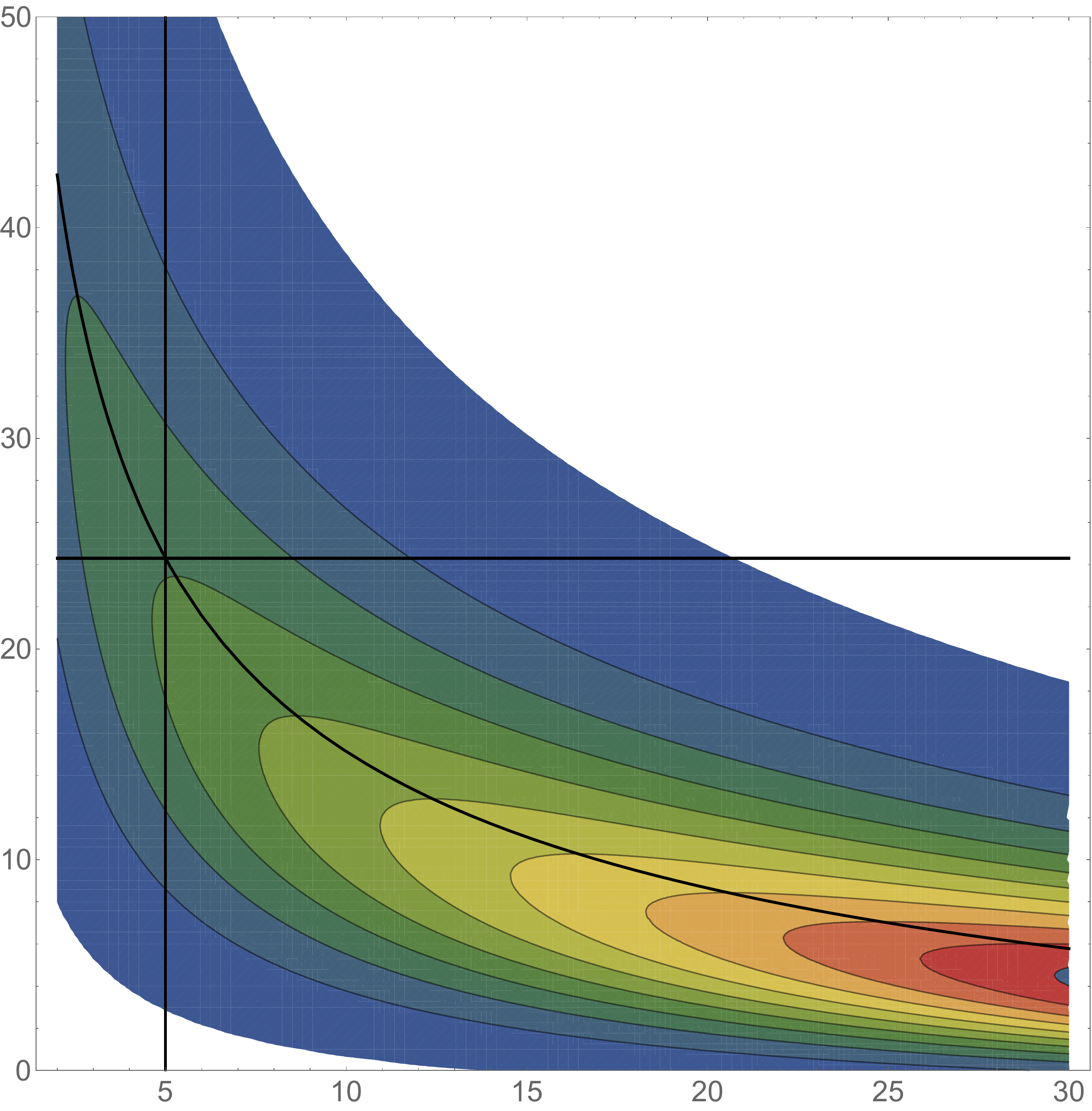}}\qquad
  \subfloat[A $3d$ plot of $p(l+m,t_i)/t$, showing only the values for the domains $A$, $B$ and $C$.  ]{\includegraphics[width=0.5\textwidth]{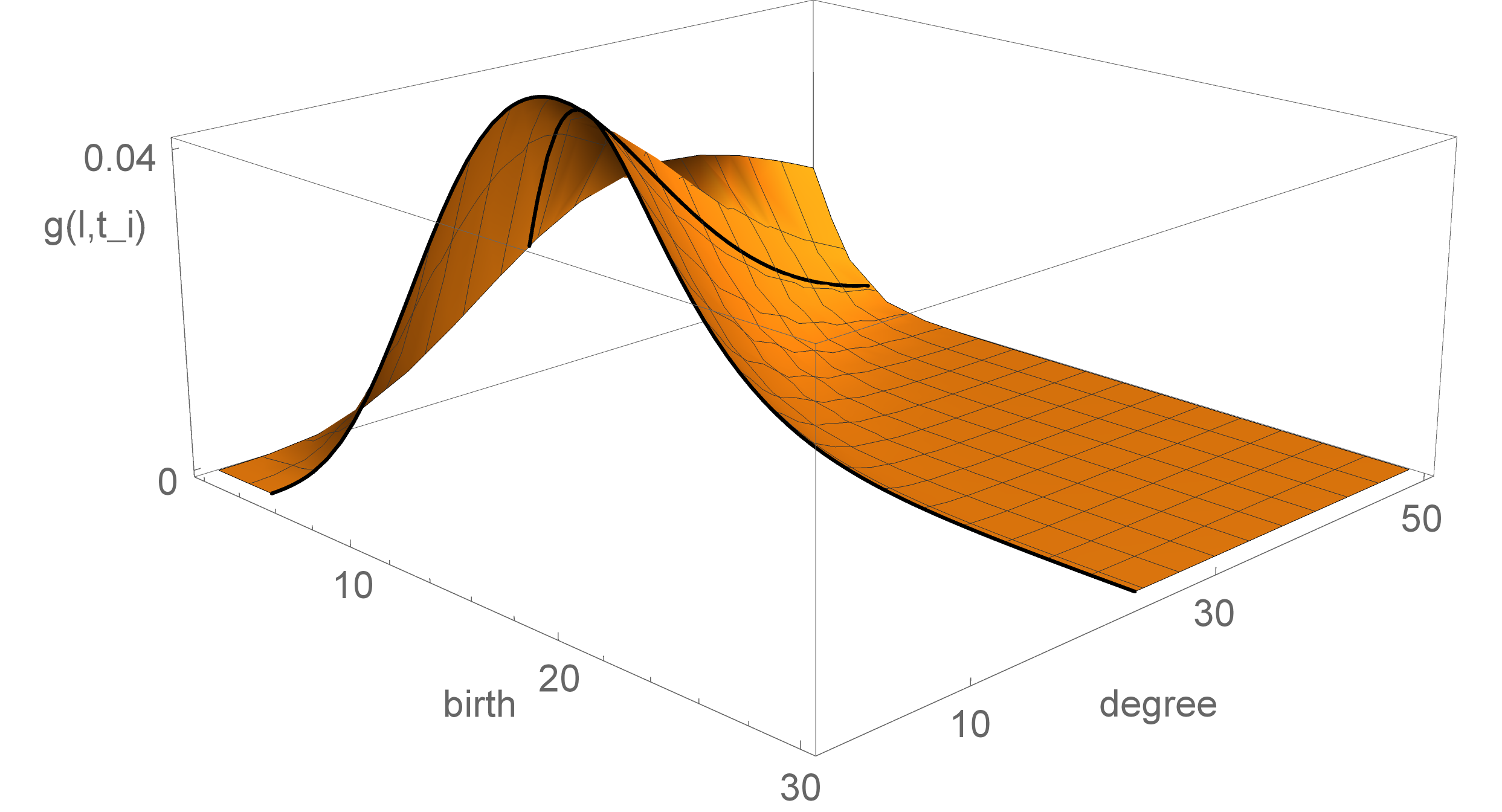}}\\
 \caption{Plots of $p(l+m,t_i)/t$. For $k<m$, $p(k,t_i)=0$, for all $t_i$. In both figures $t=100$ and $m=7$. The total probability in areas $A$ and $C$ is approximately equal. }
 \label{contour}
\end{figure}

\section{A counterexample}
\begin{figure}[htbp]
 \centering
 \subfloat[$\Tilde{P}(k;t)$ (orange), $P_{MFA}(k;t)$ (black) and the the true cumulative distribution $P(k;t)$ (blue).]{\includegraphics[width=0.45\textwidth]{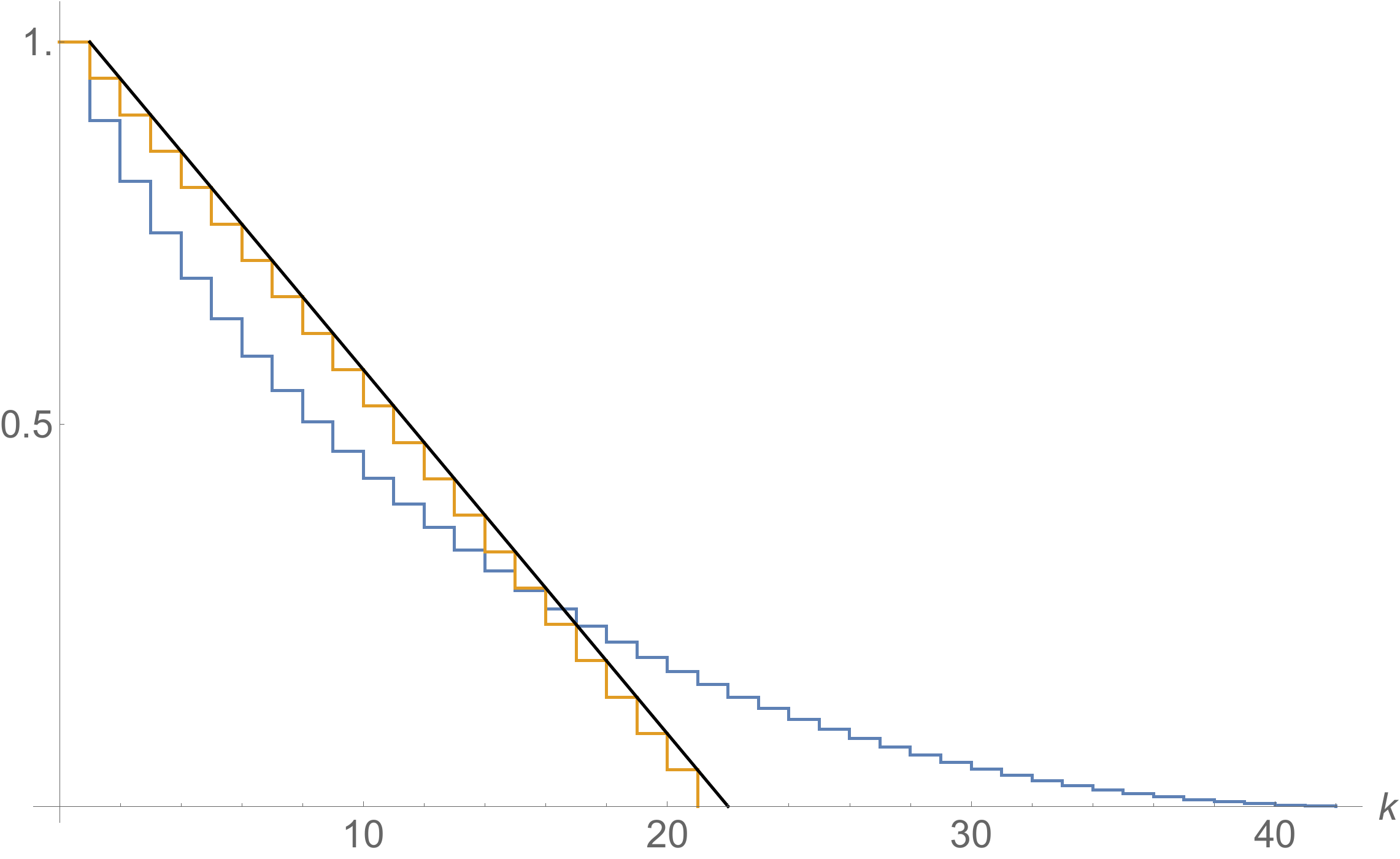}}\qquad
  \subfloat[ The approximation $p_{MFA}(k;t)$ (orange) and the true probability distribution $p(k;t)$  (blue)]{\includegraphics[width=0.45\textwidth]{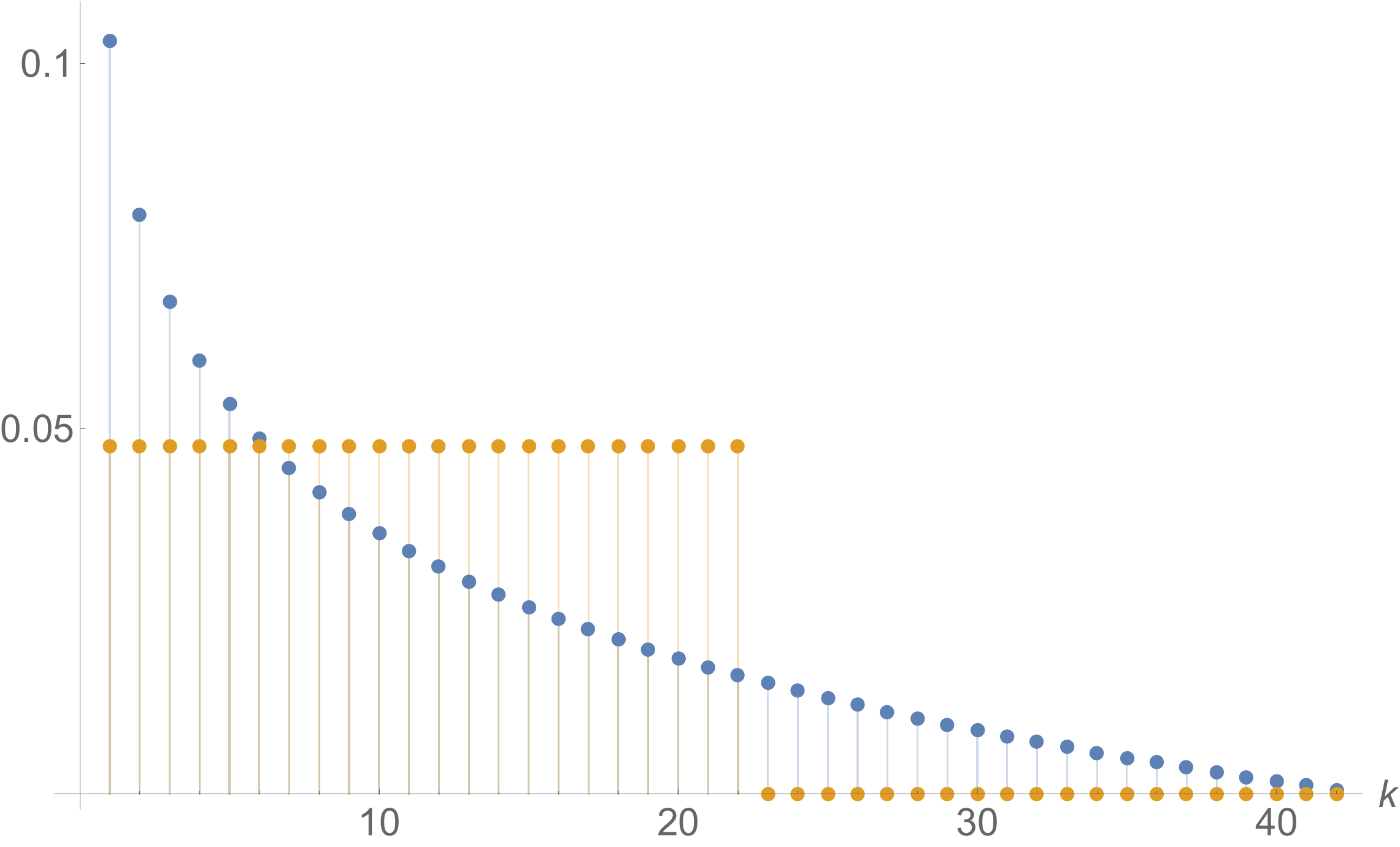}}\\
 \caption{Probability distributions corresponding to the counterexample. Here, $t=41$. }
 \label{Uni}
\end{figure}

There is nothing in the MFA that requires the network to be of the preferential attachment type. If the MFA is correct, it should work for networks with an arbitrary degree distribution.\\
Consider a random network consisting of $t$ nodes. Let $K_i(t)$ be the degree of the node $t_i$ at time $t$ and assume it has an uniform distribution over $\{1,2,\ldots,t+1-t_i\}$, for each $t_i=1, 2, \ldots, t$. The expected values of the $K_i(t)$ are
\begin{equation*}
\langle K_i(t) \rangle=(t-t_i+2)/2\, , \quad t_i=1,2,\dots, t. 
\end{equation*}
The jump points of the cumulative probability function $\Tilde{P}(k;t)$, shown in Figure \ref{Uni}, are therefore successive multiples of $1/2$:
\begin{align}
\label{uPt}
\Tilde{P}(k;t)=\begin{cases} 1 &\text{if}\, k\leq 1 \\
1-\frac{j}{t} & \text{for } j=1,2,\ldots, (t-1) \quad \text{and} \quad \frac{1+j}{2}<k\leq \frac{2+j}{2}  \\
                      0                                    & \text{otherwise \, . }     
        \end{cases} 
\end{align}
The approximation $P_{MFA}(k;t)$, which corresponds with $\Tilde{P}(k;t)$ in the jump points $k=\frac{2+j}{2}$, is given by 
\begin{align*}
P_{MFA}(k;t)=\begin{cases} 1 &\text{if}\, k\leq 1 \\
1-\frac{2}{t}(k-1)& \text{if } 1<k\leq \frac{2+t}{2}  \\
                      0                                    & \text{otherwise }     
        \end{cases} 
\end{align*}
This leads to
\begin{align*}
    p_{MFA}(k;t)=\frac{2}{t}\quad \text{for}\, k=1,2,\ldots ,(t+2)/2.
\end{align*}
The true value of $p(k;t)$ is
\begin{align*}
    p(k;t)&=\frac{1}{t}\sum_{i=1}^{t} p_i(k;t) \\
    &=\frac{1}{t}\sum_{i=1}^{t+1-k} \frac{1}{t+1-i}=\frac{1}{t}\sum_{n=k}^{t} \frac{1}{n}\\
    &\approx \frac{1}{t}\log(t/k)\, ,
\end{align*}
for large $t$ and moderately large $k$.\\
It is clear from Figure \ref{Uni} that $p_{MFA}(k;t)$ is not a good approximation of $p(k;t)$.\\

\section{Conclusion}
The MFA consists of a sequence of mathematical manipulations, resulting in the calculation of the degree distribution of a network. In the case of the BA network, its result is close to correct. Nonetheless, the MFA is not a good general method to teach students. There are many reasons for this assessment.\\
First, the method is not backed up by a proper mathematical model. This makes it impossible for students to see where approximations are made and how the MFA actually works. Students are presented with a number of steps, each of which they may remember from Calculus or Probability class, but without a model, the method can not be critically examined.\\
Second, the method is faulty. It treats the degree of a node simultaneously as deterministic and as stochastic. It changes discrete variables, such as a timestep and the degree of a node, in to continuous ones without a reason. It interpolates functions which are only defined on integers to functions defined on the reals in a haphazard way. All these fallacies lead to unresolvable contradictions.\\
Third, the mathematical model behind the MFA can be reconstructed,  and it shows that the familiar looking operations hide a lack of intuitive content. Figure \ref{MFAfig} shows that the final distribution is derived from a rather mysterious rearrangement of a different initial distribution. There is no indication why this would work.\\
Fourth, the reconstructed model contains dubious elements. It replaces a distribution, which is shown numerically to have a large variance, by a distribution concentrated in one point. Crucially, at one point an approximation is used which is completely unmotivated. This approximation works for BA networks, but only by chance. A counterexample shows that this approximation is, in general, false.\\  
Finally, a simple alternative method exists, namely the rate-equation method, which  gives the correct answer and is based on a standard mathematical technique.\\
This leads to the question why, after more than twenty years, the MFA method can be found in many textbooks and course notes, apparently having passed dozens of authors, editors and referees unscathed. Finding the answer could be an interesting exercise in the sociology of science, but ironically one of the reasons might involve preferential attachment networks themselves. De Solla Price \cite{Price} already noted that scientific papers tend to form a preferential attachment network, with references from one paper to another acting as the edges. The huge success of \cite{BarAlb}, the initial node of the citation network on preferential attachment, may have given it such an aura of authority that a critical review of its every aspect would have been considered unnecessary. 

\section*{Appendix A} 
The value of $p(k)$ can be found using the rate equation, an approximation method based on the master equation approach.\\
Let $n(k;t)$ be the expected number of nodes of degree $ k \geq m$ at time $t$. Neglecting the probability that an existing node receives two or more links from a new node, the transition probability at time $t+1$ for a node to go from $k$ links to $k+1$ links is $k/(2t)$. This leads to the following set of equations:
\begin{align*}
    n(k;t+1)&=n(k;t)+\frac{k-1}{2t}n(k-1;t)-\frac{k}{2t}n(k;t)\, , \quad k>m \\
    n(m;t+1)&=n(m;t)+1-\frac{m}{2t}n(m;t)\, .
\end{align*}
Since $p(k;t)=n(k;t)/t$, the above equations can be written as:
\begin{align*}
    2(t+1)p(k;t+1)&=2t p(k;t)+(k-1)p(k-1;t)-k p(k;t)\, , \quad k>m \\
    2(t+1)p(m;t+1)&=2t p(m;t)-m p(m;t)+2\, .
\end{align*}
Now assume that $\lim_{t \rightarrow \infty} p(k;t)=p(k)$. Making the approximation $p(k;t+1)\approx p(k;t)\approx p(k)$ for large $t$, substituting in the above equation and rearranging leads to:
\begin{align*}
    p(k)&=\frac{k-1}{k+2}p(k-1)\, , \quad k>m \\
    p(m)&=\frac{2}{m+2}\, .
\end{align*}
It is easy to check that 
\begin{align*}
    p(k)=\frac{2m(m+1)}{k(k+1)(k+2)}
\end{align*}
is the solution of this recursion.
\section*{Appendix B}
The model will be changed so that it becomes continuous in time. This model was first used in connection with BA networks in \cite{BolEtAl}.\\
Time is taken to be the finite interval $[0,T]$. New nodes are born at integer values of time. During a short time-interval $\Delta t$, at most one connection is made from the new node to an existing node. The probability that an existing node receives a connection is proportional to its degree $k$ and to $\Delta t$. The expected total number of connections made during a time-interval of unit length is $m$. This leads to the probability of attachment to a node of degree $k$ during a short time-interval $[t,t+\Delta t]$ to be $m k \Delta t/S(t) $. Here, $S(t)=2m \lfloor t \rfloor$ is the total degree of the network. For large values of $t$, the relative error in approximating $\lfloor t \rfloor$ by $t$ is small. Therefore, the probability of attachment is simplified to $k \Delta t/(2t)$. Eventually, the limit $\Delta t \rightarrow 0$ is taken. \\
The change in the model implies that its outcomes are not guaranteed to match those of the original. Numerical simulations show that the continuous-time model is a good approximation to the discrete-time one, apart for the earliest nodes. A theoretical justification can be found in \cite{BolEtAl}.\\
If the network is considered as a directed graph, the out-degree of every node is constant and equal to $m$. It is the in-degree $d$ which contains all the uncertainty. Consequently, it is useful to write the total degree as $k=d+m$. Let $D_i(t)=K_i(t)-m$ be the stochastic variable representing the in-degree, at time $t$, of the node born at $t_i$. The probability distribution for $D_i(t)$ will be denoted $q_i(d;t)$.\\ 
The master-equations for $q_i(d;t+\Delta t)$ become
\begin{align*}
q_i(0;t+\Delta t)&=(1- \frac{m\Delta t}{2t})q_i(0;t)\\
q_i(d;t+\Delta t)&=(1- \frac{(m+d)\Delta t}{2t})q_i(d;t)+  \frac{(m+d-1)\Delta t}{2t} q_i(d-1;t) \quad , \quad 1\leq d \leq T-1\\
q_i(T;t+\Delta t)&=\frac{(m+T-1)\Delta t}{2t} q_i(T-1;t)\, ,
\end{align*}
with initial conditions 
\begin{align}
\label{initial}
q_i(0;t_i)&=1\\
q_i(d;t_i)&=0\quad {\text{for all}}\,  d>0
\end{align} 
Note that there is a set of equations for each $i=1,2, \ldots, T$.\\
Rearranging the terms and taking the limit $\Delta t \rightarrow 0$ leads to the differential equations:
\begin{align}
\label{difeq2}
\frac{\text{d}}{\text{d} t}q_i(0;t)&=-\frac{m}{2t}q_i(0;t)\, , \nonumber\\
\frac{\text{d}}{\text{d} t}q_i(d;t)&=\frac{1}{2t}\left((m+d-1)q_i(d-1;t)-(m+d)q_i(d;t)\right)\quad , \quad d \geq 1 \, .
\end{align}
The equation for $q_i(0;t)$ is easily solved, and yields: 
\begin{align*}
q_i(0;t)=(\frac{t_i}{t})^{\nicefrac{m}{2}} \, ,
\end{align*}
where the initial condition (\ref{initial}) was used.\\
To derive the solution for $q_i(d;t)$ with $d \geq 1$ the following transformation is introduced: 
\begin{align*}
q_i(d;t)=\tau_i^{-(d+m)}r(d;\tau_i)\, \quad {\text{with}}\,  \tau_i=(\frac{t}{t_i})^{\nicefrac{1}{2}}\, .
\end{align*}
Substitution in equation (\ref{difeq2}) gives:
\begin{align}
\label{difeq3}
\frac{\text{d}}{\text{d} \tau_i}r(d;\tau_i)=(m+d-1)r(d-1;\tau_i)\, , \quad d \geq 1 \, .
\end{align}
After solving for $d=1$, $d=2 \ldots$, it soon becomes clear that the solution of (\ref{difeq3}) which satisfies the initial condition $r(d;1)=0$, is
\begin{align*}
r(d;\tau_i)=\binom{m+d-1}{m-1} (\tau_i-1)^d\, .
\end{align*}
This gives the result:
\begin{align}
\label{distrg}
q_i(d;t)=\tau_i^{-m}\binom{m+d-1}{m-1} (1-(\tau_i)^{-1})^d\, , \quad d=0,1,2, \ldots
\end{align}
and therefore
\begin{align}
\label{distrp}
p_i(k;t)=q_i(k-m;t)=\tau_i^{-m}\binom{k-1}{m-1} (1-(\tau_i)^{-1})^{k-m}\, , \quad k=m,m+1, \ldots
\end{align}
The function (\ref{distrg}) is known as the negative binomial distribution with parameters $r=m$ and $p=1-(\tau_i)^{-1}$$=1-(\frac{t_i}{t})^{\nicefrac{1}{2}}$.
\begin{figure}[htbp]
 \centering
 \subfloat[$t_i=2$]{\includegraphics[width=0.45\textwidth]{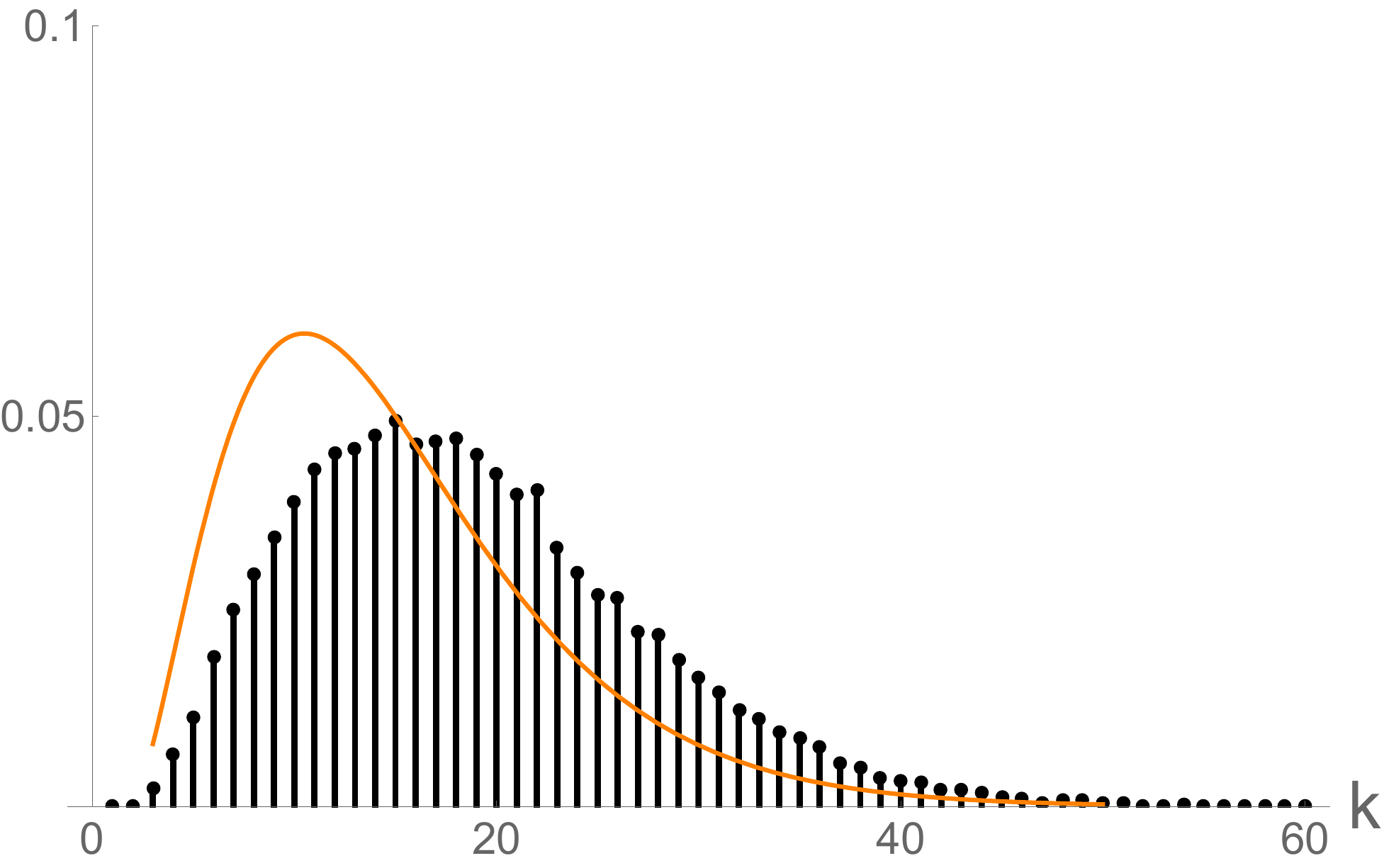}}
 \qquad
 \subfloat[$t_i=5$]{\includegraphics[width=0.45\textwidth]{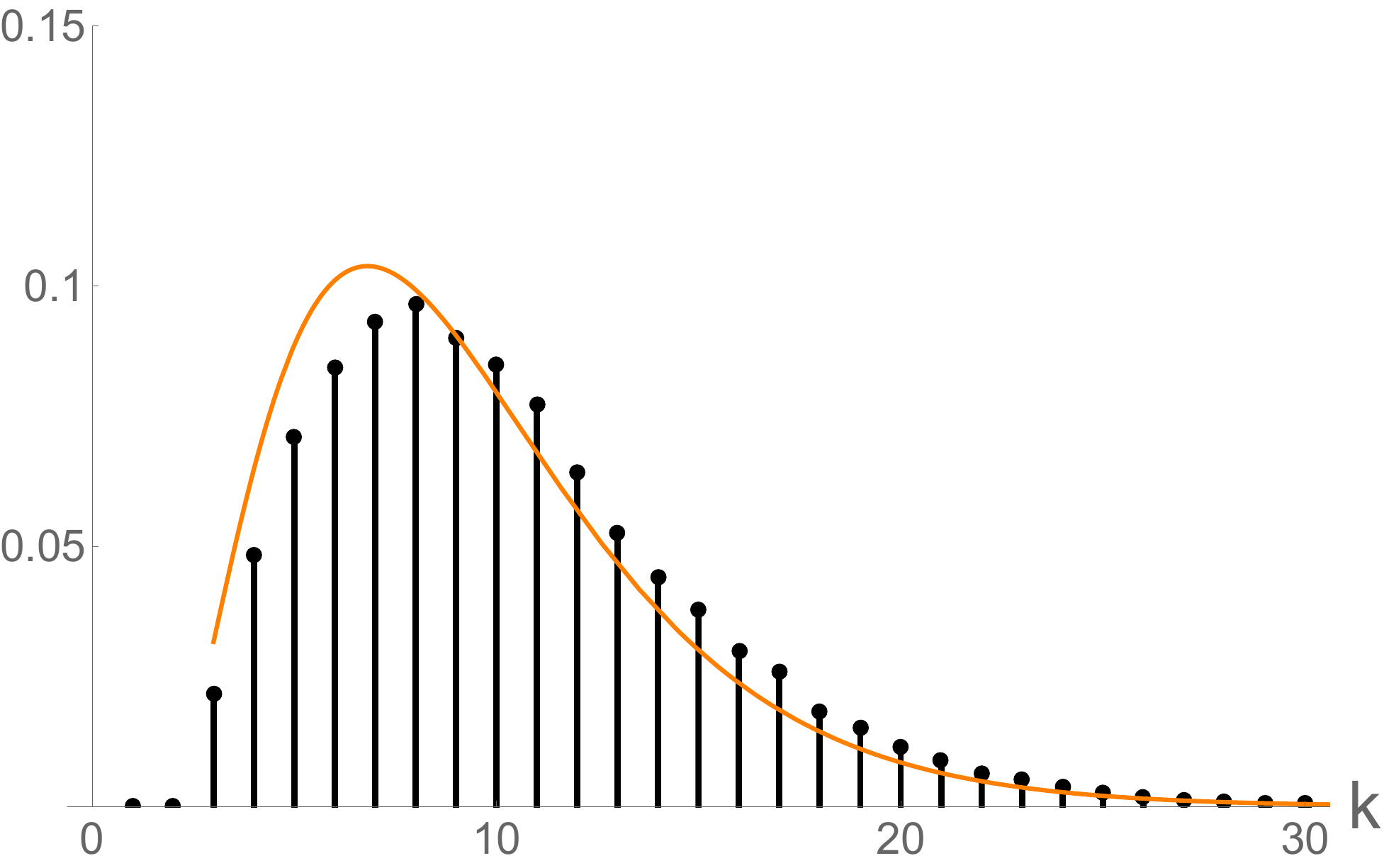}
 \label{3b}}\\
 \subfloat[$t_i=10$]{\includegraphics[width=0.45\textwidth]{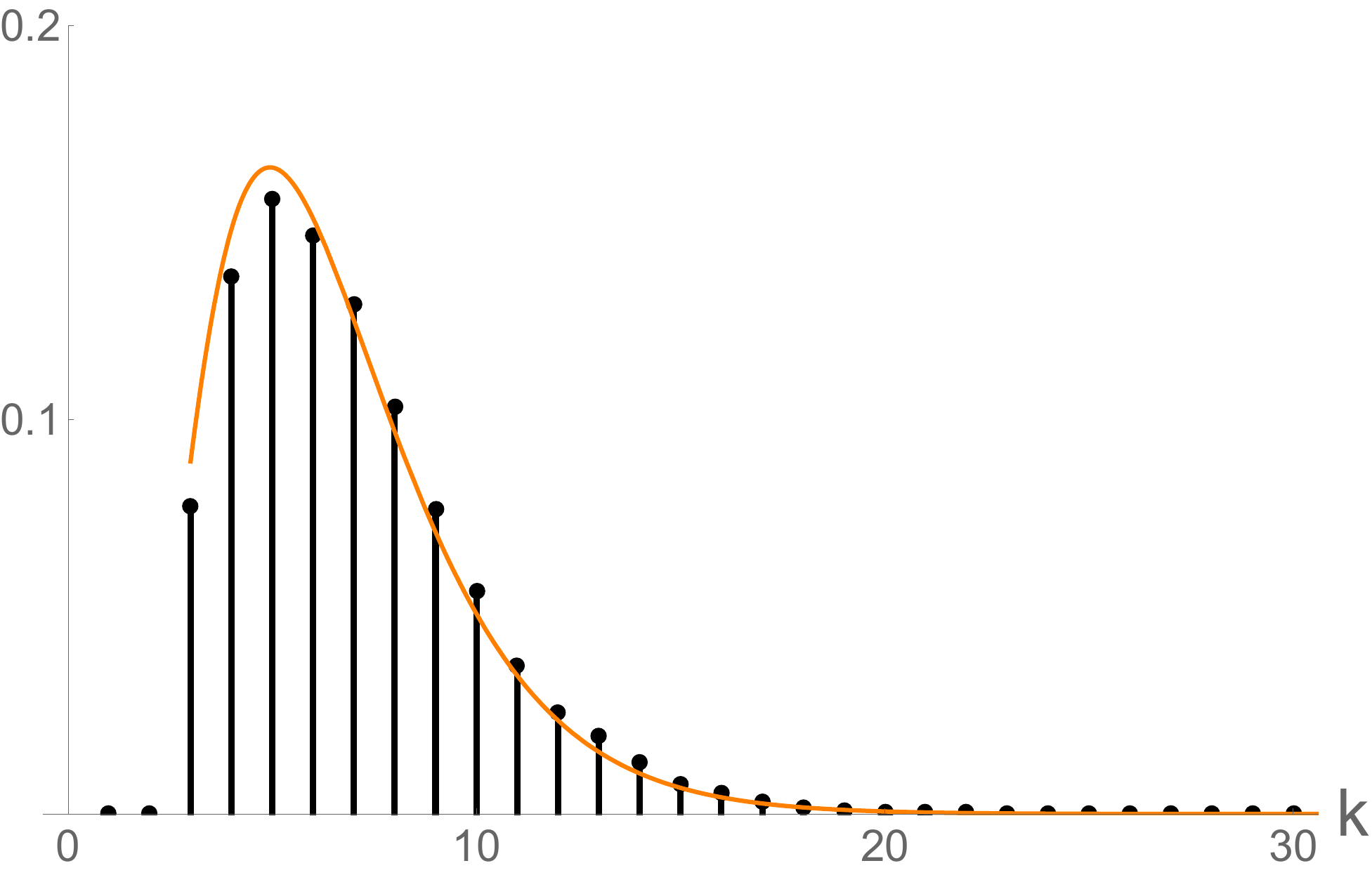}
 \label{3c}}
 \qquad
 \subfloat[$t_i=20$]{\includegraphics[width=0.45\textwidth]{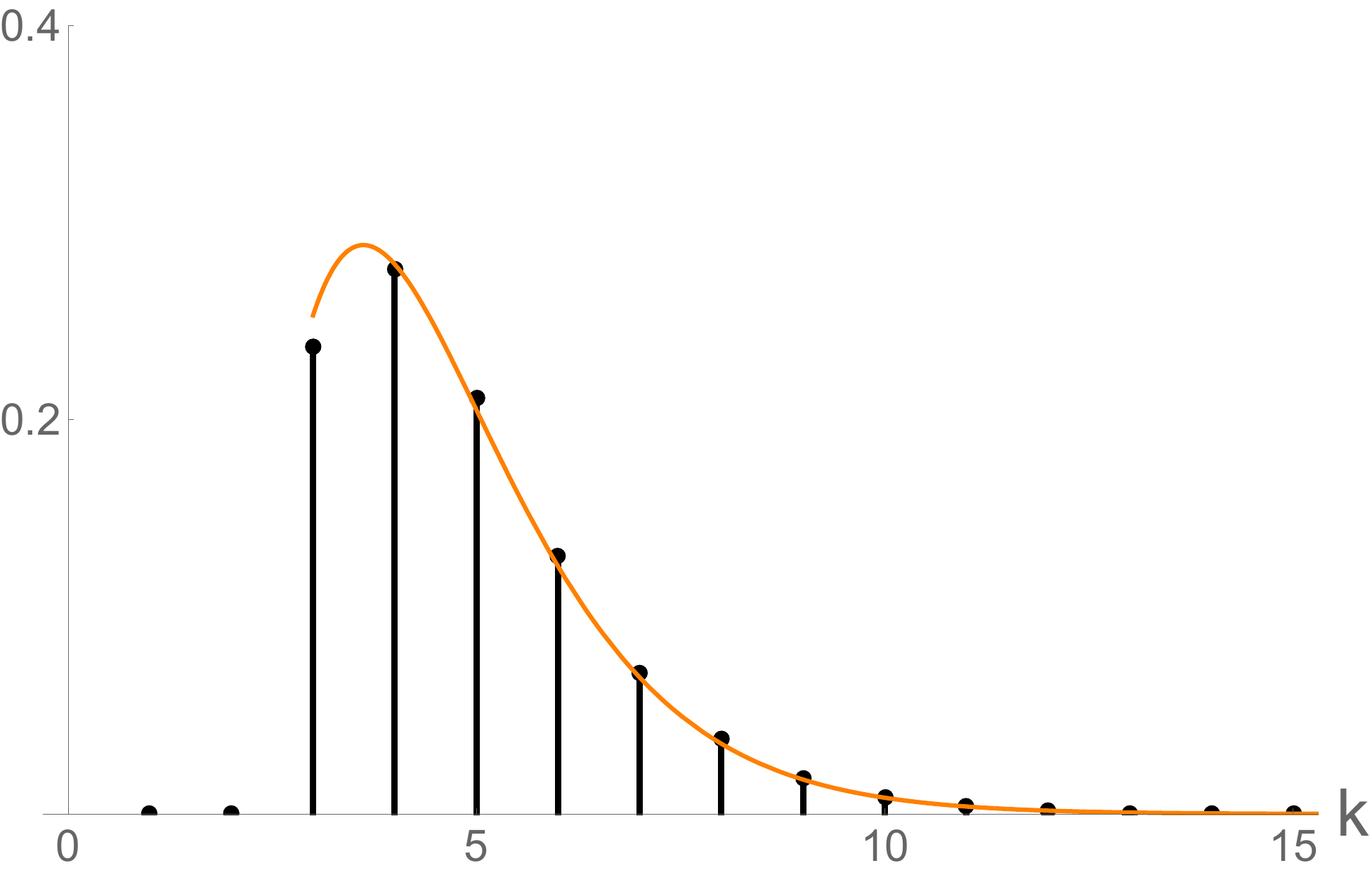}
 \label{3d}}\\
 \caption{The distribution $p_i(k;t)$ corresponding to $K_i(t)$.  In orange, the theoretical prediction $p_i(k;t)=q_i(k-m;t)$ (for clarity this function is plotted for all values of $k$, not just the integer values). In black, the numerical result based on 20000 simulations. In all graphs, $t=50$ and $m=3$.}
 \label{simuls}
\end{figure}
Figure \ref{simuls} compares simulations of the BA network to expression (\ref{distrp}). The match is very good, excepting for nodes with a very low birth date. For those nodes, the mean of the numerically calculated distribution can be appreciably larger than that of $p_i(k;t)$, although the general shape of the theoretical distribution and the simulation agree. See Figure \ref{MeanApprox} for a plot of the means of $K_i(t)$.\\
The mean and variance of the negative binomial distribution can be found in many reference sources:
\begin{align}
\langle D_i(t) \rangle&=\frac{pr}{1-p}=m (\frac{t}{t_i})^{\nicefrac{1}{2}}-m \, , \\ {\text{Var}}(D_i(t))&=\frac{pr}{(1-p)^2}=m((\frac{t}{t_i})-(\frac{t}{t_i})^{\nicefrac{1}{2}})\, .
\end{align}
Because $K_i(t)=D_i(t)+m$, the model predicts 
\begin{align}
\label{avk}
\langle K_i(t) \rangle=m (\frac{t}{t_i})^{\nicefrac{1}{2}}\, .
\end{align}
\section*{Appendix C}
\begin{figure}[htbp]
 \centering
 \includegraphics[width=0.45\textwidth]{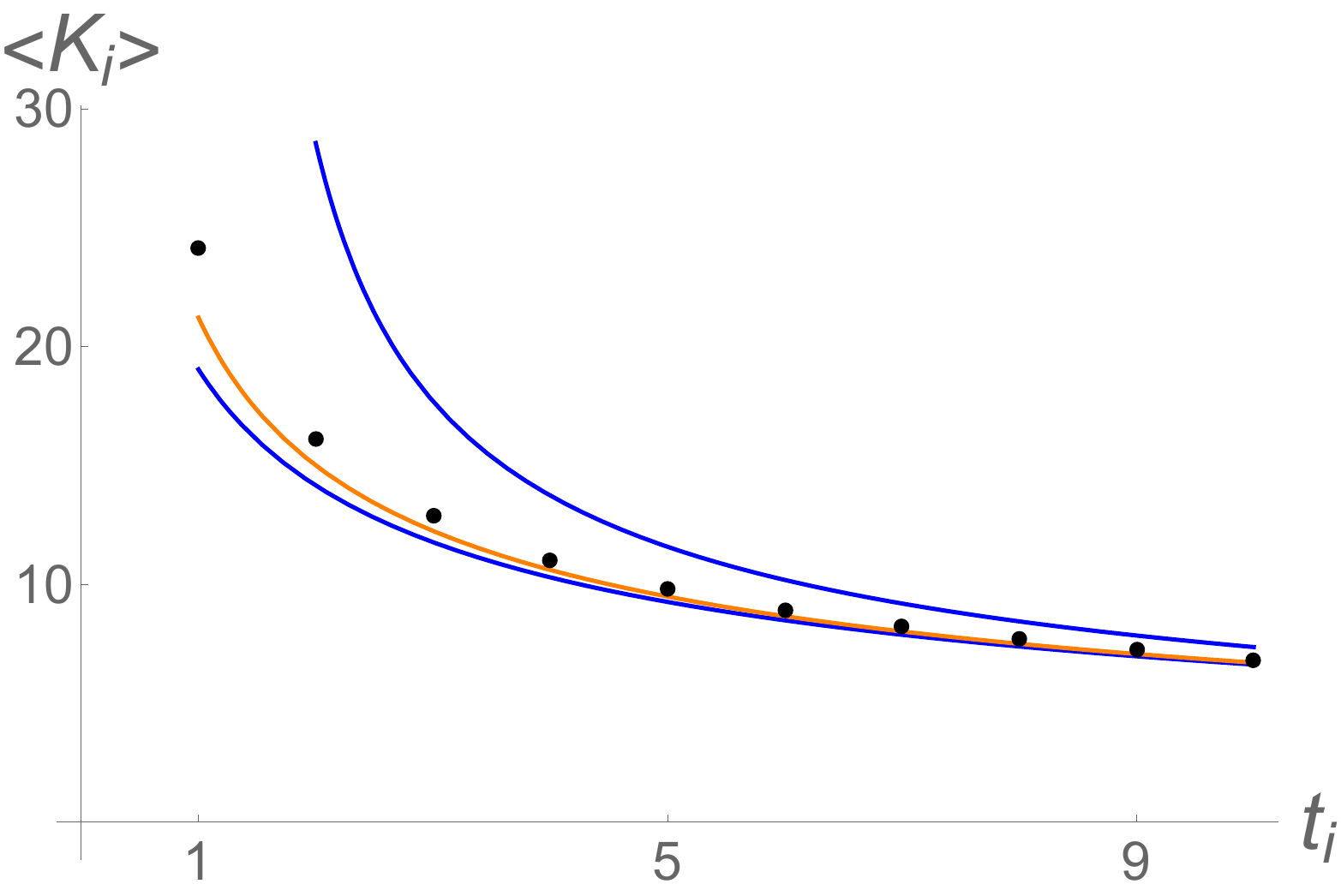}
  \caption{The blue lines are the upper and lower bounds for $ \langle K_i(t) \rangle $ using (\ref{meanK}). The orange line is $m(\frac{t}{t_i})^{\nicefrac{1}{2}}$. The black dots represent the exact value of $ \langle K_i(t) \rangle $ calculated from (\ref{SolK}). As in Figure \ref{simuls}, $t=50$ and $m=3$.   }
  \label{MeanApprox}
\end{figure}

The expected value of $K_i(t)$ can be derived by the reasoning in section \ref{method}, using discrete time steps:
\begin{align*}
  \langle K_i(t+1) \rangle  -\langle K_i(t) \rangle=\frac{\langle K_i(t) \rangle}{2t} \, , \quad \langle K_i(t_i)\rangle=m\,.
\end{align*}
This leads to the recursion
\begin{align*}
  \langle K_i(t+1) \rangle =(1+\frac{1}{2t}) \langle K_i(t) \rangle\, , \quad \langle K_i(t_i)\rangle=m\,,
\end{align*}
with solution
\begin{align}
\label{SolK}
  \langle K_i(t) \rangle =(1+\frac{1}{2(t-1)})\dots (1+\frac{1}{2t_i})m\,, \quad t>t_i\, .
\end{align}
This expression is not very informative, so an upper- and lower bound will be derived. Taking logarithms leads to 
\begin{align*}
 \log \langle K_i(t) \rangle =\sum_{k=t_i}^{t-1}\log (1+\frac{1}{2k})+\log m\,.
\end{align*}
Because $\log (1+\frac{1}{2k})$ is strictly decreasing in $k$, the sum in the right-hand side is both a Riemann lower sum and an upper sum:
\begin{align*}
\int_{t_i}^{t}\log (1+\frac{1}{2x})\, \text{d}x \leq \sum_{k=t_i}^{t-1}\log (1+\frac{1}{2k})\leq \int_{t_i-1}^{t-1}\log (1+\frac{1}{2x})\, \text{d}x\, .
\end{align*}
This leads to
\begin{align*}
 \sum_{k=t_i}^{t-1}\log (1+\frac{1}{2k})&\leq  \log(\frac{1+2(t-1)}{1+2(t_i-1)})^{\nicefrac{1}{2}}+(t-1)\log(1+\frac{1}{2(t-1)})-(t_i-1)\log(1+\frac{1}{2(t_i-1)})\\
 \sum_{k=t_i}^{t-1}\log (1+\frac{1}{2k})&\geq  \log(\frac{1+2t}{1+2t_i})^{\nicefrac{1}{2}}+t\log(1+\frac{1}{2t})-t_i\log(1+\frac{1}{2t_i}) \, .
\end{align*}
For large $t$, the terms $(t-1)\log(1+\frac{1}{2(t-1)})$ and $t\log(1+\frac{1}{2t})$ can both be replaced by $\frac{1}{2}$ and the terms $t-1$ and $t+1$ both by $t$.\\
Taking exponents again yields the bounds:
\begin{align}
\label{meanK}
  m(\frac{e\,t}{t_i+1/2})^{\nicefrac{1}{2}}(1+\frac{1}{2t_i})^{-t_i} \leq \langle K_i(t) \rangle\leq 
     m(\frac{e\,t}{t_i+1/2})^{\nicefrac{1}{2}}(1+\frac{1}{2(t_i-1)})^{-(t_i-1)}\, .
 \end{align}
 When $t_i$ is moderately large, both the upper bound and the lower bound in expression (\ref{meanK})  are close to the value $m(\frac{t}{t_i})^{\nicefrac{1}{2}}$ found in (\ref{avk}). For small values of $t_i$, calculating $ \langle K_i(t) \rangle $ using (\ref{SolK}) shows that for all but the smallest values of $t_i$, $m(\frac{t}{t_i})^{\nicefrac{1}{2}}$ is also an excellent approximation for $ \langle K_i(t) \rangle $, see Figure \ref{MeanApprox}. 
 \section*{Acknowledgements}
 I thank Rob Bisseling, Anton van de Ven and Theo Ruijgrok for their comments. Th. R. also wrote some of the code for the simulations.

\end{document}